\documentclass[twocolumn,preprintnumbers,aps,pra,superscriptaddress,amsmath,amssymb]{revtex4-1}

\usepackage{graphicx}
\usepackage{dcolumn}
\usepackage{bm}
\usepackage{bbm}
\usepackage{hyperref}
\usepackage{color}
\usepackage{multirow}
\usepackage{physics}

\begin{document}

\title{Magnetic noise from ultra-thin abrasively deposited materials on diamond}

\author{Scott E. Lillie}
\affiliation{School of Physics, The University of Melbourne, VIC 3010, Australia}
\affiliation{Centre for Quantum Computation and Communication Technology, School of Physics, The University of Melbourne, VIC 3010, Australia}

\author{David A. Broadway}
\affiliation{School of Physics, The University of Melbourne, VIC 3010, Australia}
\affiliation{Centre for Quantum Computation and Communication Technology, School of Physics, The University of Melbourne, VIC 3010, Australia}

\author{Nikolai Dontschuk}
\affiliation{School of Physics, The University of Melbourne, VIC 3010, Australia}
\affiliation{Centre for Quantum Computation and Communication Technology, School of Physics, The University of Melbourne, VIC 3010, Australia}

\author{Ali Zavabeti}
\affiliation{School of Engineering, RMIT University, VIC 3001, Australia}

\author{David A. Simpson}
\affiliation{School of Physics, The University of Melbourne, VIC 3010, Australia}

\author{Tokuyuki Teraji}
\affiliation{National Institute for Materials Science, Tsukuba, Ibaraki 305-0044, Japan}

\author{Torben Daeneke}
\affiliation{School of Engineering, RMIT University, VIC 3001, Australia}

\author{Lloyd C. L. Hollenberg}
\email{lloydch@unimelb.edu.au}
\affiliation{School of Physics, The University of Melbourne, VIC 3010, Australia}
\affiliation{Centre for Quantum Computation and Communication Technology, School of Physics, The University of Melbourne, VIC 3010, Australia}

\author{Jean-Philippe Tetienne} 
\email{jtetienne@unimelb.edu.au}
\affiliation{School of Physics, The University of Melbourne, VIC 3010, Australia}

\date{\today}
	
\begin{abstract}
	
Sensing techniques based on the negatively charged nitrogen-vacancy (NV) centre in diamond have emerged as promising candidates to characterise ultra-thin and $2$D materials. An outstanding challenge to this goal is isolating the contribution of $2$D materials from undesired contributions arising from surface contamination, and changes to the diamond surface induced by the sample or transfer process. Here we report on such a scenario, in which the abrasive deposition of trace amounts of materials onto a diamond gives rise to a previously unreported source of magnetic noise. By deliberately scratching the diamond surface with macroscopic blocks of various metals (Fe, Cu, Cr, Au), we are able to form ultra-thin structures (i.e. with thicknesses down to $<1$\,nm), and find that these structures give rise to a broadband source of noise. Explanation for these effects are discussed, including spin and charge noise native to the sample and/or induced by sample-surface interactions, and indirect effects, where the deposited material affects the charge stability and magnetic environment of the sensing layer. This work illustrates the high sensitivity of NV noise spectroscopy to ultra-thin materials down to sub-nm regimes -- a key step towards the study of $2$D electronic systems -- and highlights the need to passivate the diamond surface for future sensing applications in ultra-thin and 2D materials.

\end{abstract}

\maketitle

In recent years, quantum sensing techniques utilising negatively charged nitrogen-vacancy (NV) centres in diamond have been turned to interrogate condensed matter systems external to the diamond host \cite{Casola2018}. Among these, ultra-thin and $2$D materials are of keen interest, given the unique consequences of low-dimensionality for electronic~\cite{Cao2018a}, magnetic \cite{Gong2017,Huang2017}, and transport properties \cite{Cao2018}; and one that NV sensing is particularly applicable to, given its high-sensitivity at nm length scales. Experiments using single NV centres have demonstrated nuclear magnetic resonance measurements of atomically thin materials \cite{Lovchinsky2017}, while dense NV ensemble measurements have magnetically imaged charge transport in mono-layer graphene~\cite{Tetienne2017}. Additionally, the ability to discriminate between different transport regimes and identify impurities within $2$D materials based on their magnetic noise profiles has been outlined theoretically \cite{Argawal2017}, and should be similarly applicable to $1$D systems \cite{Rodriguez-Nieva2018}. 

\begin{figure}[t]
	\begin{center}
		\includegraphics[width=0.95\columnwidth]{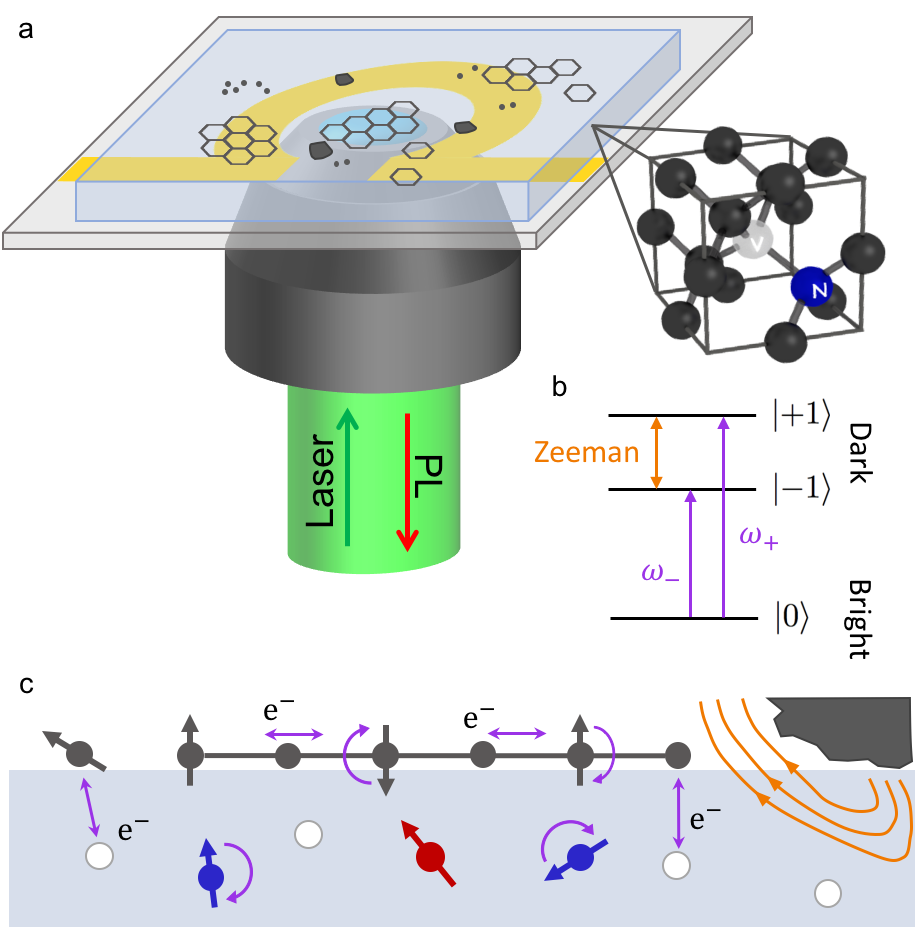}
		\caption{\textbf{a} Schematic of the quantum imaging of $2$D and nano-scale materials. Samples are transferred onto a single crystal diamond (blue) which contains a dense layer of near-surface NV centres (unit cell). Optical readout is achieved by wide-field illumination with a $532$\,nm laser, and the photoluminescence (PL) is collected on a sCMOS camera for imaging. \textbf{b} Ground state structure of the spin-$1$ NV centre, where the Zeeman splitting of the dark $\ket{\pm1}$ states, relative to the bright $\ket{0}$ state, allows static magnetic fields to be imaged. The transition frequencies $\omega_\pm$ determine the spectral sensitivity of the NV spin-state to magnetic noise, and hence their relaxation rates. \textbf{c} Illustration of potential magnetic noise sources (purple) in $2$D magnetic imaging with near-surface NV centres (red). Spin- and charge-noise may arise from the $2$D material of interest (bonded structure), or from atomic and particulate inclusions, which may also exude static magnetic fields (orange). Magnetic noise may also arise from defects within the diamond, such as substitutional nitrogen (blue), or vacancies (white), the population and dynamics of which may be affected by the sample at the surface.}
		\label{Fig0}
	\end{center}
\end{figure}

Quantum imaging of $2$D materials with NV-diamond substrates presents a set of challenges, primarily regarding sample-induced changes to the diamond surface and sensing layer, that may obscure direct contributions from the sample, which itself may be altered during fabrication or transfer processes, or by sample-surface interactions \cite{Dev2014,Lippert2016}. Furthermore, there exists a range of possible contaminants present in the standard preparation techniques used for NV quantum sensing, the effects of which have not been well described, particularly in the context of $2$D materials. In this work, we utilise a soft abrasion technique to deliberately deposit trace amounts of material onto diamond substrates, and characterise the resulting magnetic noise via quantum measurements of the near-surface ensemble of NV spins. Initially, results are presented for the abrasive deposition of inox steel, which is shown to leave ferromagnetic nano-particles on the diamond down to $20$\,nm in size, the magnetic fields of which are imaged by optically detected magnetic resonance (ODMR). Additionally, a thin coverage of the material, down to $<1$\,nm in thickness, is measured in the scratched regions, and found to be the source of broadband magnetic noise that quenches the NV-layer spin-lattice relaxation time, $T_1$. Similar results are found from the deposition of other metallic materials, namely copper, chromium, and gold, whereas insulating materials, such as silicon, produce no such effect.

The negatively charged nitrogen-vacancy (NV) centre in diamond is an atom-sized defect comprised of a substitutional nitrogen and adjacent vacancy [Fig.~\ref{Fig0}a], that is capable of measuring static-to-GHz frequency magnetic fields with great sensitivity \cite{Doherty2013,Rondin2014}. The spin-state dependent photoluminescence (PL) of this spin-$1$ system [Fig.~\ref{Fig0}b] allows all-optical readout of the NV spin-state, and hence dense NV ensembles can be used for wide-field magnetic imaging of samples at the diamond surface [Fig.~\ref{Fig0}a]. In these experiments, we use a $\langle100\rangle$ oriented single crystal diamond, with surfaces as-grown by chemical vapour deposition \cite{Teraji2015}, and implanted with nitrogen ions at $4$ and $6$ keV, giving NV centres at mean depths of about $10$\,nm and $15$\,nm respectively \cite{Tetienne2018}. All NV measurements were performed in ambient conditions using a purpose-built wide-field microscope similar to that described in Ref. \cite{Simpson2016}, with a spatial resolution close to the diffraction limit of $\approx300$\,nm. In this configuration, the NV-layer is sensitive to magnetic noise arising from $2$D and ultra-thin materials at the surface, either in the form of spin-noise or charge-noise (i.e. magnetic noise associated with the movement of charge carriers), and also noise arising from defect states within the diamond, the population and dynamics of which may be altered by the presence of the sample \cite{Tetienne2018}. These scenarios are illustrated in Fig.~\ref{Fig0}c.

\begin{figure}[t]
	\begin{center}
		\includegraphics[width=0.5\textwidth]{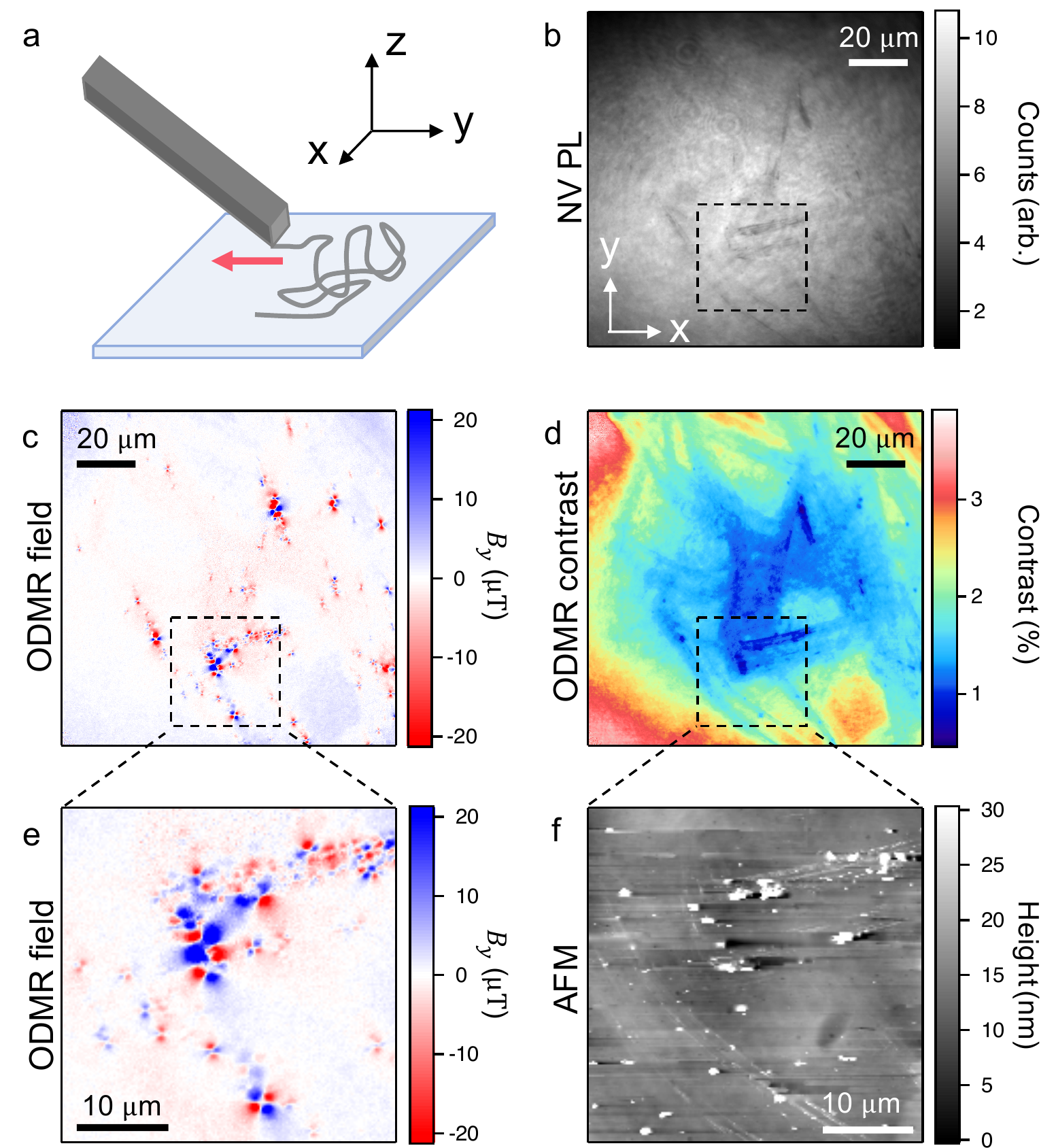}
		\caption{\textbf{a} Schematic showing abrasive deposition of materials on diamond. The material (grey) is dragged across the diamond surface (blue), leaving material residues down to sub-nm thicknesses. \textbf{b} PL image acquired from the NV-layer, showing streaked PL quenching due to the deposited inox steel. \textbf{c} Magnetic field image ($B_y$ component) as reconstructed from an ODMR measurement of the NV layer. The z-scale is capped at $\pm20$\,$\mu$T to highlight lesser field strengths. The largest fields measured are approximately $100$\,$\mu$T in magnitude. \textbf{d} ODMR contrast map for a single NV family spin transition (see appendix B, Fig. 7). \textbf{e} and \textbf{f} show a zoom-in of the static field image and an AFM topography map respectively, of the same region, showing the correlation of static magnetic fields with the location of particles of size $>20$\,nm in height. In \textbf{f}, the z-scale is capped at $30$\,nm to emphasise smaller features. Central particles are $>$\,$500$\,nm in height. The regions presented in \textbf{b}, \textbf{c}, and \textbf{d} are identical.}
		\label{Fig1}
	\end{center}
\end{figure}

The abrasive deposition of materials onto diamond was achieved by manually dragging the tip of a macroscopic block of a given material across the fixed diamond surface [Fig.~\ref{Fig1}a], in a region previously characterised by NV measurement (appendix \ref{BackgroundCharac}). First, we explore the consequences of depositing inox steel by this technique, a particularly relevant study given its ubiquity in laboratory environments in many forms, such as tweezers, which were used for the deposition in this instance. Figure \ref{Fig1}b is an image of the deposition region, taken by collecting the NV-layer PL after excitation with a $532$\,nm laser. The image shows quenching of the NV PL by up to $20$\% in streaked patterns, $10$'s of $\mu$m in length. The observed quenching is possibly due to a number of effects, including a F\"orster resonant energy transfer (FRET) effect \cite{Tisler2011,Tisler2013}, a reduction in local laser intensity due to scattering from the sample, changes to the lifetime and angular distribution of the NV emission \cite{Buchler2005}, or discharging of the NV centres (becoming charge neutral, NV$^0$  \cite{Hauf2011}) induced by the sample. An optical micrograph of this region shows micron-scale particulate features within the regions of greater PL quenching (appendix \ref{BackgroundCharac}). We associate both of these features with the abrasive deposition, having previously characterised the same region with both PL and optical microscopy.

To image the static magnetic fields in this region, an ODMR measurement was performed, with an external magnetic field applied across the NV-layer, such that the two electron-spin transition frequencies of each of the four NV orientation families are resolved \cite{Tetienne2017,Tetienne2018c}. The spin-state dependent PL of the NV centre allows these transition frequencies to be measured optically, and acquisition of the PL with a camera allows pixel-by-pixel ODMR spectra to be constructed. Fitting the spectra in accordance with the NV spin-Hamiltonian, static magnetic fields can be mapped across the field of view and calibrated to the lab-frame coordinates \cite{Tetienne2017,Tetienne2018c}. Figure \ref{Fig1}c maps the magnetic field strength oriented along the $y$-direction, $B_y$, which is representative of the in-plane magnetic field (the $B_x$ and $B_z$ components are shown in appendix \ref{FieldArtefact}). The regions identified previously by their strong PL quenching contain ferromagnetic moments, with field strengths measured up to $\sim100$\,$\mu$T. Comparing a zoomed-in magnetic field image [Fig.~\ref{Fig1}e] with an atomic force microscopy (AFM) image of the same region [Fig.~\ref{Fig1}f], it is clear that the magnetic dipole signatures are correlated with particles present at the diamond surface, with heights ranging from $20$\,nm to $500$\,nm. Energy-dispersive X-ray spectroscopy (EDXS) of larger particles from a similarly prepared sample confirm the presence of iron and oxygen, suggesting that the particles are iron oxide, Fe$_2$O$_3$/Fe$_3$O$_4$ (appendix \ref{EDXS}).

In addition to reconstructing magnetic field strengths from the NV spin-transition frequencies, the optical contrast of these transitions can be mapped across the same region. Figure \ref{Fig1}d shows a map of the ODMR contrast for the lowest frequency transition, where the contrast is significantly reduced at sites of ferromagnetism, due to broadening from the static fields \cite{Tetienne2018c}. However, the contrast map shows additional reductions in contrast beyond the regions of large magnetic field strengths and PL quenching, suggesting an additional effect of the deposited material on the NV-layer. The AFM data shows that one such region is bounded by $<10$\,nm tall ridges oriented along the direction of the abrasions, which do not show in the optical, PL, or magnetic micro-graphs. Given the transition of iron oxide from ferromagnetism to super-paramagnetism at sizes $<20$\,nm \cite{Li2017}, imaging fluctuating magnetic fields is necessary to further characterise the system.

\begin{figure*}[t]
	\begin{center}
		\includegraphics[width=1.0\textwidth]{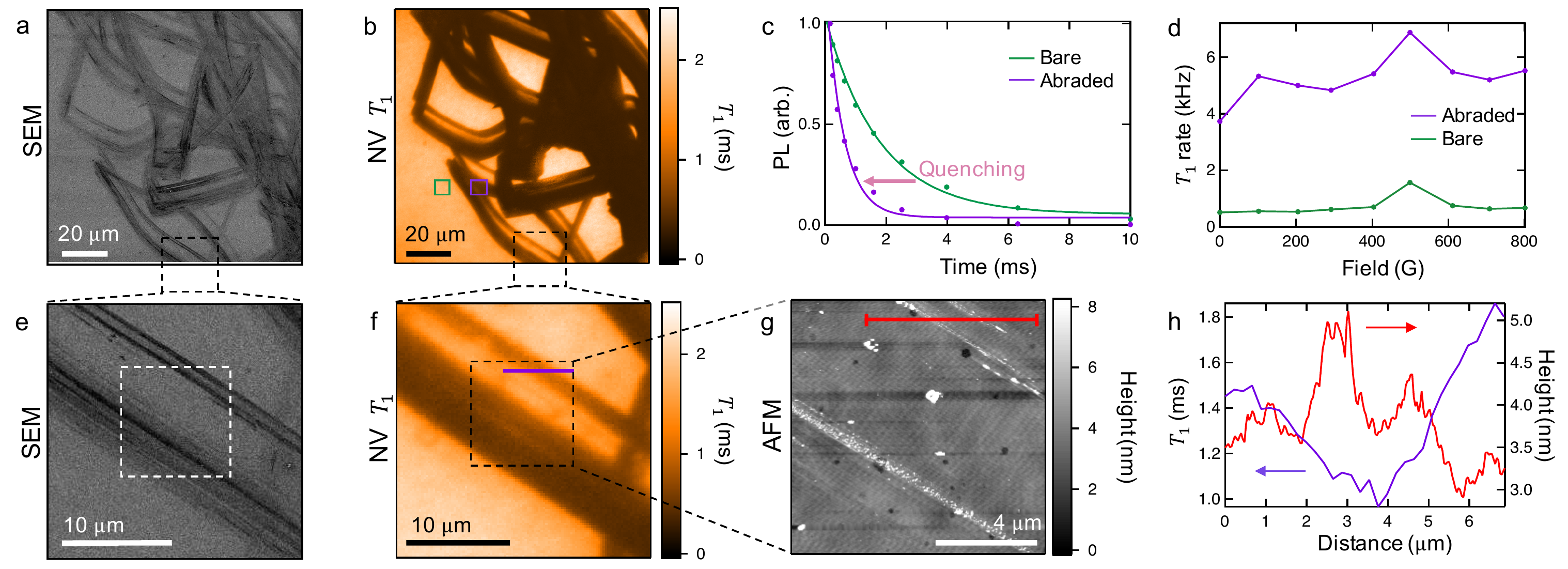}
		\caption{\textbf{a} SEM and \textbf{b} NV $T_1$-relaxation images showing significant quenching of $T_1$-relaxation time under material residues left by abrasive deposition, in the same region as shown in Fig.~\ref{Fig1}b - d. \textbf{c} Individual $T_1$-relaxation curves from the areas highlighted in \textbf{b}. The relaxation time has been reduced from $1.79$\,ms in the untouched region (green) to $0.57$\,ms under the deposition (purple). \textbf{d} $T_1$-relaxation rate ($1/T_1$) spectra of an untouched region (green) and a region under the material deposition (purple). \textbf{e} and \textbf{f} show zoomed SEM and $T_1$-relaxation images, respectively, of the highlighted regions. \textbf{g} AFM topography of the region highlighted in \textbf{e} and \textbf{f}. The z-scale is capped at $8$\,nm to emphasise the smaller features. The tallest particles within the aggregate are $\sim20$\,nm in height. \textbf{h} AFM (red) and $T_1$ (purple) taken along the lines shown in \textbf{f} and \textbf{g}. Approximately $1$\,nm step heights are seen across the edge of parallel bands, which define the quenched $T_1$ region. The $<1$\,$\mu$m separation of the bands is at the resolution of the $T_1$ imaging.}
		\label{Fig2}
	\end{center}
\end{figure*}

$T_1$-relaxometry \cite{Cole2009} is an established technique in NV-sensing which allows the detection of magnetic noise at the electron spin sub-level transition frequency \cite{Kaufmann2013,Tetienne2013,Ermakova2013,Steinert2013,Sushkov2014}. In this measurement, the NV is initialised in the bright state, $\ket{0}$, by a polarising laser pulse, and left to evolve towards a thermal mixture with the dark states, $\ket{\pm1}$, before being optically read out by a second laser pulse. The observed decay in readout PL for this measurement is normalised by that from an intercalated sequence, which includes a microwave $\pi$-pulse, resonant with $\omega_\pm$, before the readout laser pulse, such that common mode variations are removed from the measurement. Here, all $T_1$-relaxometry imaging were performed at zero-field, i.e. with the NV spin-transition frequency at $\omega_\pm=2.87$\,GHz, optimising acquisition time and optical contrast, unless otherwise stated.

Figure \ref{Fig2}b shows a $T_1$-relaxometry image of the previously analysed region of abrasively deposited inox steel. Here, we see up to an order of magnitude reduction in the $T_1$-relaxation time of the NV-layer under the abraded region, which covers a far greater area than suggested by the PL and static magnetic field images. Comparing this data to a scanning electron microscope (SEM) image of the same region, we see the same pattern in the secondary electron emission contrast [Fig.~\ref{Fig2}a]. Individual relaxation curves used to compose the $T_1$ map are shown in Fig.~\ref{Fig2}c, for neighboring areas beside and under the abrasion, with $T_1$-relaxation times of $1.79$\,ms and $0.57$\,ms respectively. This corresponds to an increase in relaxation rate $\Delta\Gamma = (1/T_1)_{\rm abraded}-(1/T_1)_{\rm bare}\approx1.8$\,kHz. Additionally, the noise spectra of a bare and more heavily abraded diamond regions is shown in figure \ref{Fig2}d, measured by aligning the external magnetic field with a single NV orientation and varying the field strength, using the NV $\ket{0}$ $\leftrightarrow$ $\ket{-1}$ transition as a spectral filter \cite{Hall2016,Wood2016}. Over a range of $800$ G, there is an approximately constant offset of $4.5$\,kHz between the spectra of the bare diamond and the abraded region, suggesting that the abrasive deposition adds a broadband source of magnetic noise to the sample. The enhanced relaxation rate around $512$ G in both spectra is due to a cross-relaxation resonance with unpaired electron spins intrinsic to the diamond \cite{Hall2016,Tetienne2018}.

One striking difference between the SEM and $T_1$ images presented is the clear texturing of the abraded region seen in SEM, as compared to the more uniform $T_1$ maps. This difference is emphasised by comparing higher-resolution SEM images with a zoom-in of the $T_1$ map [Fig.~\ref{Fig2}e and f respectively]. The optical resolution limited $T_1$ map shows relatively flat quenching under the abraded regions, where the SEM image shows large variations in contrast, and clear streaking along the direction of the abrasion. The topography of this region, as mapped by AFM, shows some of the same structures [Fig.~\ref{Fig2}g]. The dark streak running along the edge of the abraded region is a collection of nano-particles, $5$\,-\,$10$\,nm in height. An approximately $1.0$\,nm step height is seen across the upper branch of deposited material, that matches spatially with the observed $T_1$ quenching within the resolution of our imaging [Fig.~\ref{Fig2}h]. The definition of the lower branch in AFM and $T_1$ is less well correlated across its full width, suggesting a sparse deposition below the noise floor of the AFM is responsible for the quenching here.

\begin{figure*}[t]
	\begin{center}
		\includegraphics[width=1.0\textwidth]{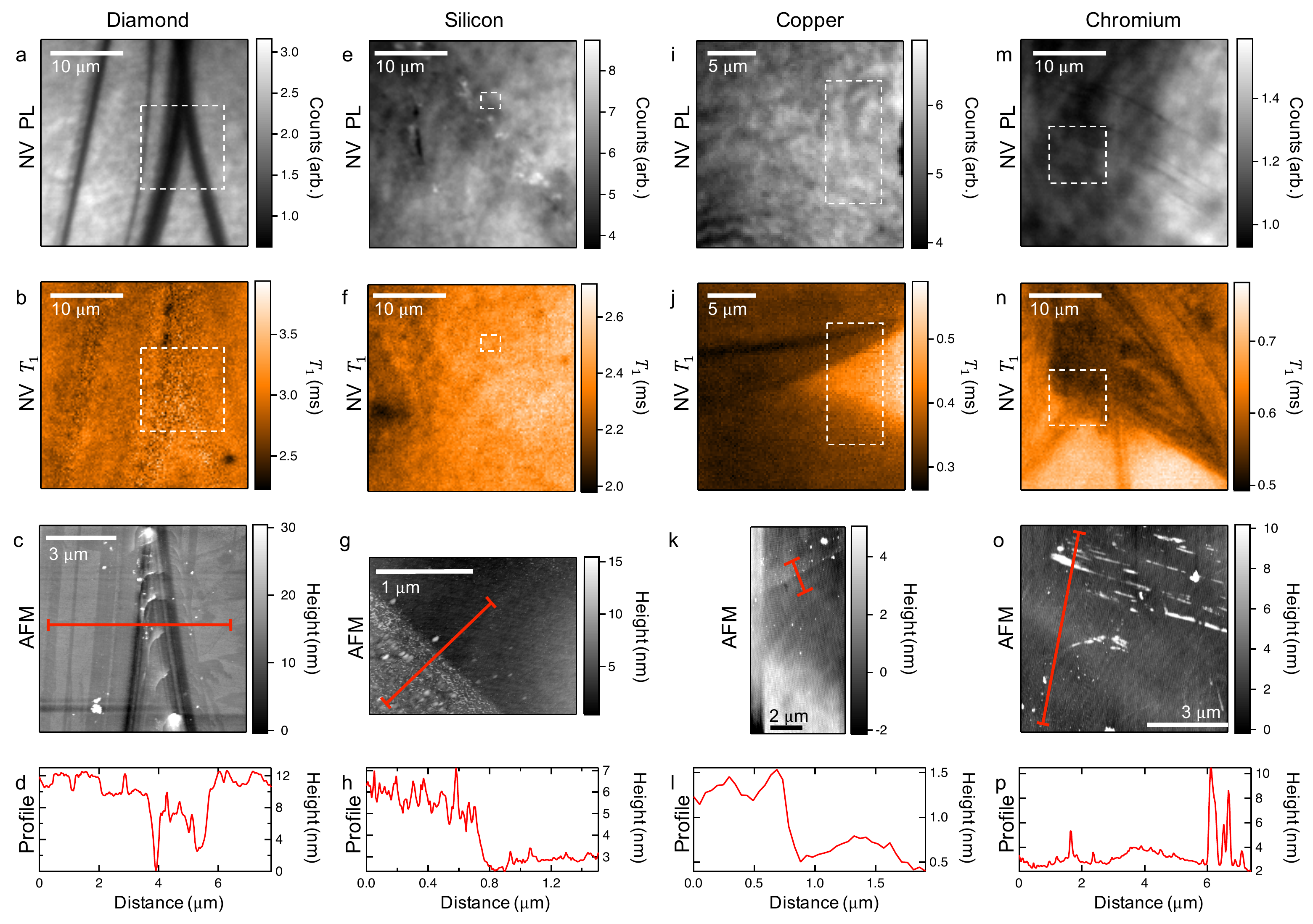}
		\caption{\textbf{a}, \textbf{b}, \textbf{c}, and \textbf{d} PL, $T_1$-relaxometry, AFM images, and AFM profile, respectively, of a region abraded with diamond scribe. The location of the AFM images region is indicated in the PL and $T_1$ images (white), and the AFM profile location is shown in the AFM image (red). \textbf{e} - \textbf{h}, \textbf{i} - \textbf{l}, and \textbf{m} - \textbf{p} As previously stated, but shown for regions of abrasively deposited silicon, copper, and chromium respectively.}
		\label{Fig3}
	\end{center}
\end{figure*}

To gain further insight into the origin of this magnetic noise, we repeated the previous experiments of soft abrasion deposition, replacing the inox steel (from tweezers) with a piece of silicon  (from a wafer), and blocks of pure copper and chromium. Additionally, in a control experiment, a diamond scribe was used instead to study the possibility of structural changes due to the abrasion process. Figure \ref{Fig3} presents PL, $T_1$-relaxometry, and AFM images of the four additional materials studied. The abrasion of NV-diamond substrate with a diamond scribe results in a dramatic quenching of NV PL under the abraded regions [Fig.~\ref{Fig3}a], however, the $T_1$-relaxometry map of the same region, shows no significant changes [Fig.~\ref{Fig3}b]. The data here is expectedly noisy due to the reduced optical readout. AFM imaging of these PL features show that the abrasion has, predictably, chipped the diamond surface by up to $10$\,nm for the strongly quenched regions, with more shallow $2$\,nm cavities running parallel to the deeper trench [Fig.~\ref{Fig3}c and d]. Given the initial NV depth profile of $0-20$\,nm, strong PL quenching from the $10$\,nm deep cavitation is explained by both the removal of some NV centres, and the impacted charge stability of those remaining, which are brought closer to the surface \cite{Hauf2011,Yamano2017}. The constant $T_1$-relaxation across the field of view is commensurate with studies of dense NV ensembles, where the $T_1$ is limited by the noise within the NV-layer rather than that at the surface \cite{Tetienne2018}, and demonstrates that the enhanced magnetic noise with inox steel does not arise from a simple removal of material at the diamond surface. 

The data for the abrasive deposition of silicon onto the diamond surface is shown in Figs.~\ref{Fig3}e - h. Figure \ref{Fig3}e shows a PL quenching under the abraded region of up to $15$\%, likely due to the mechanisms discussed in relation to the inox steel deposition. Comparing the Si PL image to the $T_1$-relaxation map of the same region [Fig.~\ref{Fig3}f], it is clear that there is no $T_1$ quenching due to the deposited Si, despite the approximately $3$\,nm thick coverage of the diamond, as shown by AFM [Fig.~\ref{Fig3}g and h].

The results for copper and chromium [Figs. \ref{Fig3}i - l and m - p respectively], however, bare resemblance to those for inox steel. The abrasive deposition of copper, despite not resulting in a PL quenching [Fig.~\ref{Fig3}i], does quench $T_1$-relaxation [Fig.~\ref{Fig3}j]. Imaging the same region with AFM [Fig.~\ref{Fig3}k and l], a $1$\,nm step height from the bare diamond to the abrasively deposited copper is associated with a $T_1$ quenching from $0.4$\,ms to $0.3$\,ms, i.e. an increase to the NV relaxation rate of $\Delta\Gamma\approx3$\,kHz. Chromium, on the other hand, shows some streaks of quenched PL [Fig.~\ref{Fig3}m], that run parallel to a large $T_1$ quenched feature [Fig.~\ref{Fig3}n]. AFM imaging of the same region shows that the PL quenched features are due to some apparently metallic deposition, $10$\,-\,$20$\,nm in height [Fig.~\ref{Fig3}o and p], however, there is no clear step height correlated with the $T_1$ quenched regions, despite a $0.5$\,kHz increase to the relaxation rate in the abraded region. Again, this suggests a sparse coverage of material, close to the noise floor of the AFM ($\sim0.5$\,nm), is responsible for the enhanced noise. Similar results are shown for the abrasive deposition a copper-beryllium composite and gold, as well as repeated depositions of copper and inox steel (appendix \ref{Materials}).

The data presented so far has demonstrated that the abrasive deposition of three metallic materials, inox steel, copper, and chromium, each result in a quenching of $T_1$-relaxation time of the NV-layer, whereas two semi-conducting materials, diamond and silicon, give no such effect. We now move to discuss the potential mechanisms by which the NV $T_1$ time could be quenched by abrasive depositions of metallic materials. The $T_1$-relaxation time of solid state spin qubits is sensitive to both phonon activity and magnetic noise at the qubit transition frequency \cite{Jarmola2012}. Here, we exclude the possibility of phonon-mediated quenching, given the negligible impact that the deposition has upon the diamond thermal vibrations, and discuss the magnetic contributions in terms of spin-noise and charge-noise.

Magnetic noise arising from fast-fluctuating spin-states in close proximity to the NV-layer has been previously observed to quench $T_1$ relaxation times \cite{Kaufmann2013,Tetienne2013,Ermakova2013,Steinert2013,Sushkov2014,Rosskopf2014}. The data presented for inox steel is consistent with such an explanation, where the large coverage of nano-particles, likely iron oxide, below a threshold size ($\sim20$\,nm), exists in a superparamagnetic state, and hence quench the NV-layer $T_1$ under these regions. These findings are commensurate with previous studies of Fe$_2$O$_3$ and Fe$_3$O$_4$ at room temperature \cite{Schmid-Lorch2015,Tetienne2016,Li2017}. Similar $T_1$ effects have been observed in NV sensing from modest densities of electron spins in various paramagnetic species \cite{Kaufmann2013,Tetienne2013,Ermakova2013,Steinert2013,Sushkov2014,Pelliccione2014,Simpson2017}. Therefore, a possible explanation for the $T_1$ quenching observed from the other materials studied (Cu, Cr, Au) is that they reside at the diamond surface in some paramagnetic form \cite{Schultz1965,Kindo1990,Gao2010,Banobre-Lopez2003}, which may be induced by interactions with the surface \cite{MaMari2015}.

The metallic nature of the bulk materials whose abrasive depositions quench $T_1$-relaxation times warrants a discussion of the role played by their conductivity. Recent work in NV sensing has demonstrated enhanced $T_1$-relaxation due to charge-noise, namely Johnson-Nyquist noise arising from metallic films; an effect which scales with the conductivity of the deposited metals \cite{Kolkowitz2015,Ariyaratne2017}. These experiments, however, studied metallic films at a minimum thickness of $60$\,nm, due to reduced conductivity at thicknesses less than the electron mean free path \cite{Ariyaratne2017,Zhang2004,Lacy2011}. To demonstrate that quenching due to Johnson-Nyquist noise is negligible in our system for this reason, we entertain the idea that abrasive deposition results in a $\sim1$\,nm thin metallic film coverage, and use the model developed in Ref. \cite{Ariyaratne2017} to calculate the resulting $T_1$ quenching. Accounting for reduced conductivity in sub-mean free path thin film thicknesses, as outlined in Ref. \cite{Lacy2011}, we find that the calculated noise from continuous films thicker than our depositions is insufficient to explain the observed quenching for all materials studied (appendix \ref{JohnsonCalc}). Additionally, the finite lateral extent of our particulate coverage would similarly reduce the conductivity, further discounting the possibility of $T_1$ quenching by Johnson-Nyquist noise \cite{Ariyaratne2017}.

A final possible explanation for the observed $T_1$ quenching, is by an indirect effect, where the deposited material alters the dynamics and filling of existing charge traps at the diamond surface and within the NV-layer. Reduced $T_1$ relaxation times due to thermally fluctuating low density surface states have been observed in single NV samples \cite{Rosskopf2014}, and the strong dependence of $T_1$ times of dense near-surface NV ensembles, similar to those used in this work, on surface treatments has been noted \cite{Tetienne2018} (with possibly related effects seen in bulk NV samples~\cite{Giri2018,Manson2018}). For these dense NV-ensembles, charge trap densities within the sensing layer can be as high as one per (nm)$^2$, corresponding to $10$ charge traps created for each implanted nitrogen \cite{DeOliveira2017}, and hence, addition of surface donors at a similar density, could substantially change the filling and dynamics of these state. The low area density of donors needed to achieve this would also explain why a perfect correlation of sample topography and $T_1$ quenching is not seen, as the density required to quench $T_1$ relaxation may be far below AFM noise floor. Imaging unpaired electron spins by double electron-electron resonance measurements is suggestive of such an indirect effect on the defects states associated with the diamond (appendix \ref{DEER}). 

In this work, we have demonstrated the application of multi-modal NV sensing to ultra-thin systems arising from the abrasive deposition of a range of materials on the diamond substrate. These systems were analysed using PL, ODMR, and $T_1$ imaging techniques, finding that the deposition of bulk metallic materials quenches the NV-layer $T_1$ relaxation, under sub-nm thick layers of material. Explanations for this effect have been discussed, including direct spin-noise and charge-noise arising from the sample, which in general cannot be separated from indirect effects that alter the population and dynamics of defects intrinsic to the diamond. This highlights an outstanding problem for NV sensing of $2$D materials, namely, the need to passivate the diamond surface such that intrinsic sample properties are not conflated with induced changes to the sensing environment. Progress has been made in this area, by reducing the number of additional defects in the sensing layer \cite{DeOliveira2017}, however, further solutions may involve targeted surface functionalisation \cite{Ryan2018} or the addition of a capping layer to fix the interface chemistry close to the sensing layer. Including an electron reservoir, such that the filling of intrinsic diamond defects is robust against sample effects, is another potential solution. Success in this direction will enhance NV imaging technology for characterising the expanding library of $2$D materials accessible to fabrication \cite{Zavabeti2017,Zhou2018}.

We thank Liam Hall and Alastair Stacey for useful discussions. This work was supported by the Australian Research Council (ARC) through grants CE110001027, FL130100119, DE170100129. J.-P.T. acknowledges support from the University of Melbourne through an Early Career Researcher Grant. D.A.B. and S.E.L. are supported by an Australian Government Research Training Program Scholarship. T.T. acknowledges the support of Grants-in-Aid for Scientific Research from the Ministry of Education, Culture, Sports, Science, and Technology, Japan (No. 15H03980, 26220903, 16H03861, and 16H06326), and a Japan Science and Technology Agency (JST) CREST Grant Number JPMJCR1773, Japan.

\appendix
 
\section{Background characterisation and optical imaging} \label{BackgroundCharac}
 
\begin{figure}[t]
 	\begin{center}
 		\includegraphics[width=0.5\textwidth]{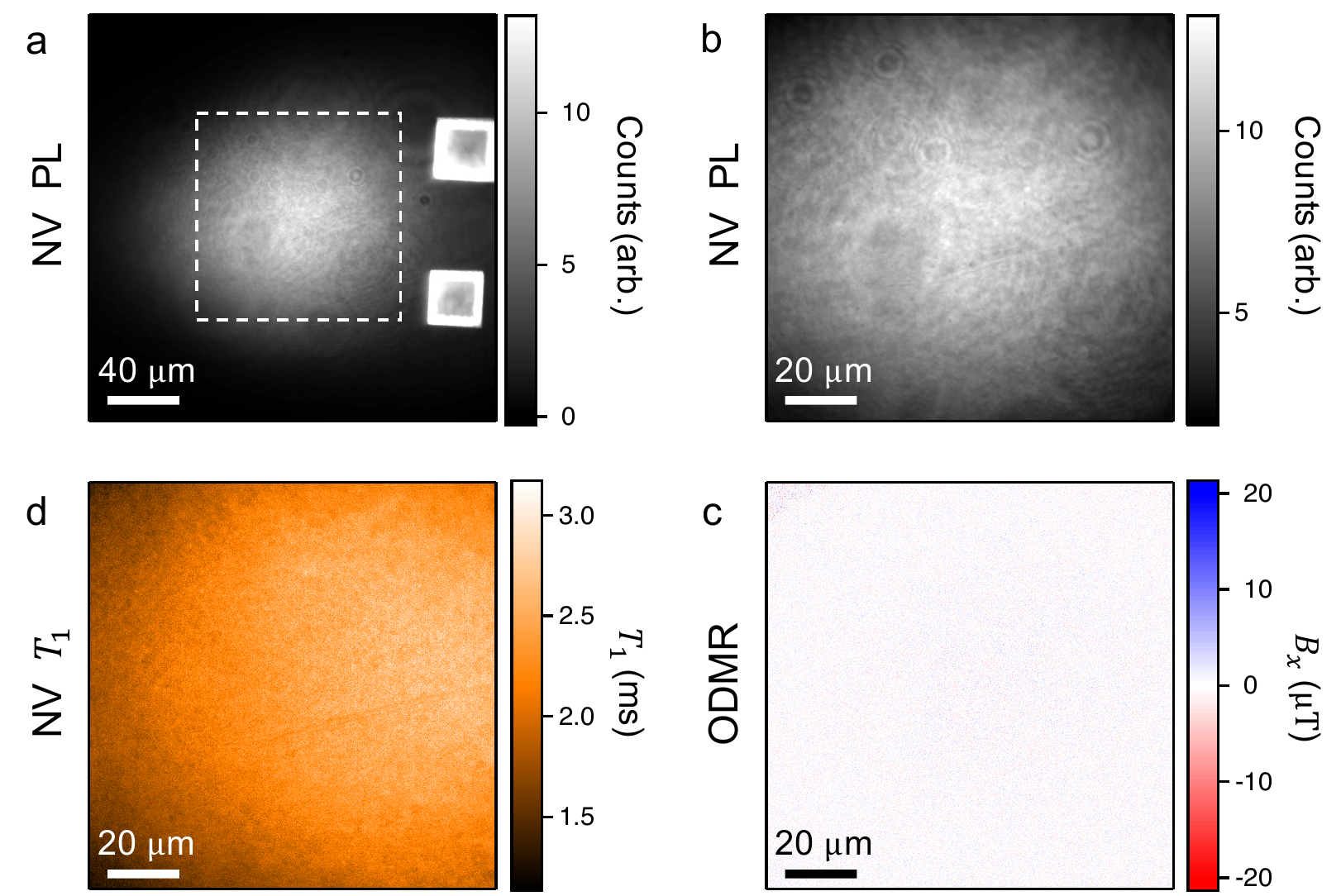}
 		\caption{\textbf{a} PL image of the extended area surrounding the region in which inox steel was deposited (white box). The bright diamond pyramids were used to reliably align the image, and compare data before and after deposition. \textbf{b} Background PL characterisation of the target region for inox steel deposition. \textbf{c} Background static field map, $B_x$, of same region. \textbf{d} Background $T_1$-relaxation map showing a slight modulation, correlated with the laser illumination, of the otherwise clean region.}
 		\label{FigS1}
 	\end{center}
\end{figure}

Prior to the abrasive deposition of the materials studied in this work, a thorough background characterisation was undertaken for each targeted area. All ensemble NV diamonds used in this work were cleaned by a Bristol boil technique (boiling mixture of sulfuric acid and sodium nitrate), prior to background characterisation and abrasive deposition. The areas characterised were chosen due to their close proximity to pyramid structures in the diamond surface, arising from the growth procedure, which served as markers by which the regions of interest could be located [Fig. \ref{FigS1}a]. The background PL, $T_1$, and static field images for the region later scratched with inox steel, are shown in Figs. \ref{FigS1}b-d. Each of the background images are free of the features highlighted in post-scratch figures in the main text, indicating that the identified features can be attributed to the abrasive deposition. A slight modulation of the background $T_1$ that is correlated with the laser illumination is observed, which we attribute to an imperfect NV spin initialisation, and highlights potential issues in interpreting $T_1$ features strongly correlated with PL. Similar background characterisations were undertaken for the regions deposited with the other materials exhibited in the paper. 
 
\begin{figure}[t]
 	\begin{center}
 		\includegraphics[width=0.5\textwidth]{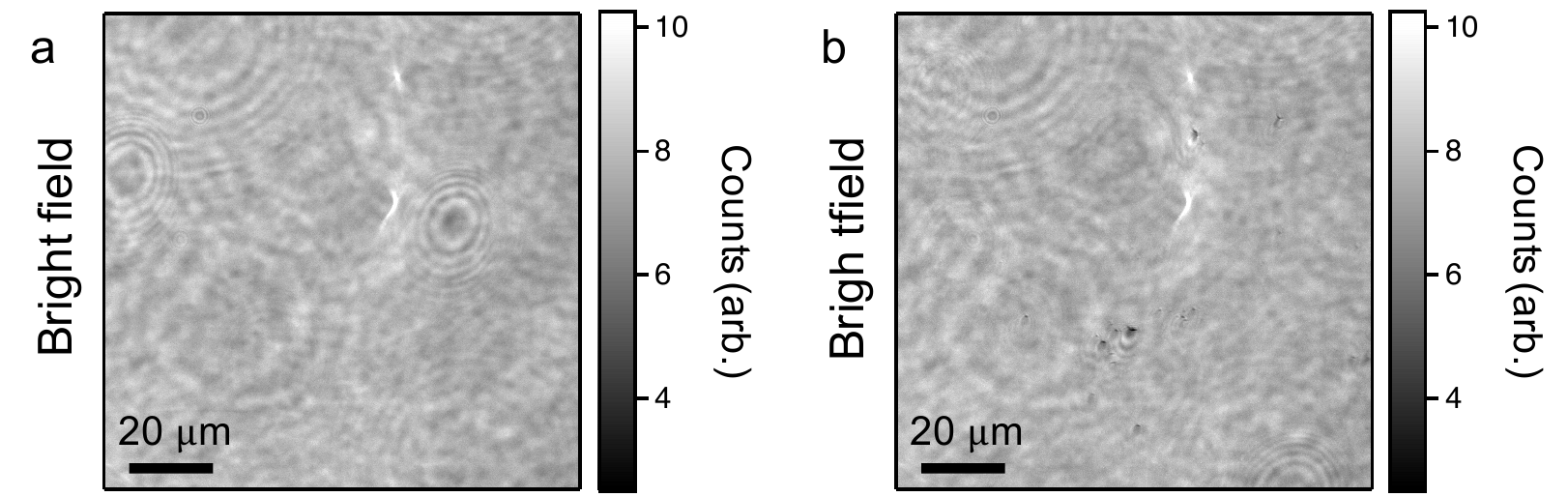}
 		\caption{Optical micrographs of the inox steel region before, \textbf{a}, and after, \textbf{b}, the abrasive deposition. Some dark shadows are seen in regions associated with regions seen to exhibit great PL quenching.}
 		\label{FigS2}
 	\end{center}
\end{figure}

In addition the NV based imaging presented in the main text, standard bright-field optical microscopy of the regions of interest was performed, by illuminating the sample with a simple lamp. Figure \ref{FigS2}a and b shows optical micrographs of the inox steel region of interest before and after the abrasive deposition. The images are practically identical with the exception of a collection of dark shadows cast in regions located within areas of greater PL quenching, which were associated with particulate inclusions approaching micron scales. The lack of optical contrast from the deposited layer was the first suggestion that the deposited layer is indeed thin, as was later confirmed by AFM imaging.
 
\section{Static magnetic field mapping by ODMR} \label{FieldArtefact}
 
\begin{figure}[t]
 	\begin{center}
 		\includegraphics[width=0.5\textwidth]{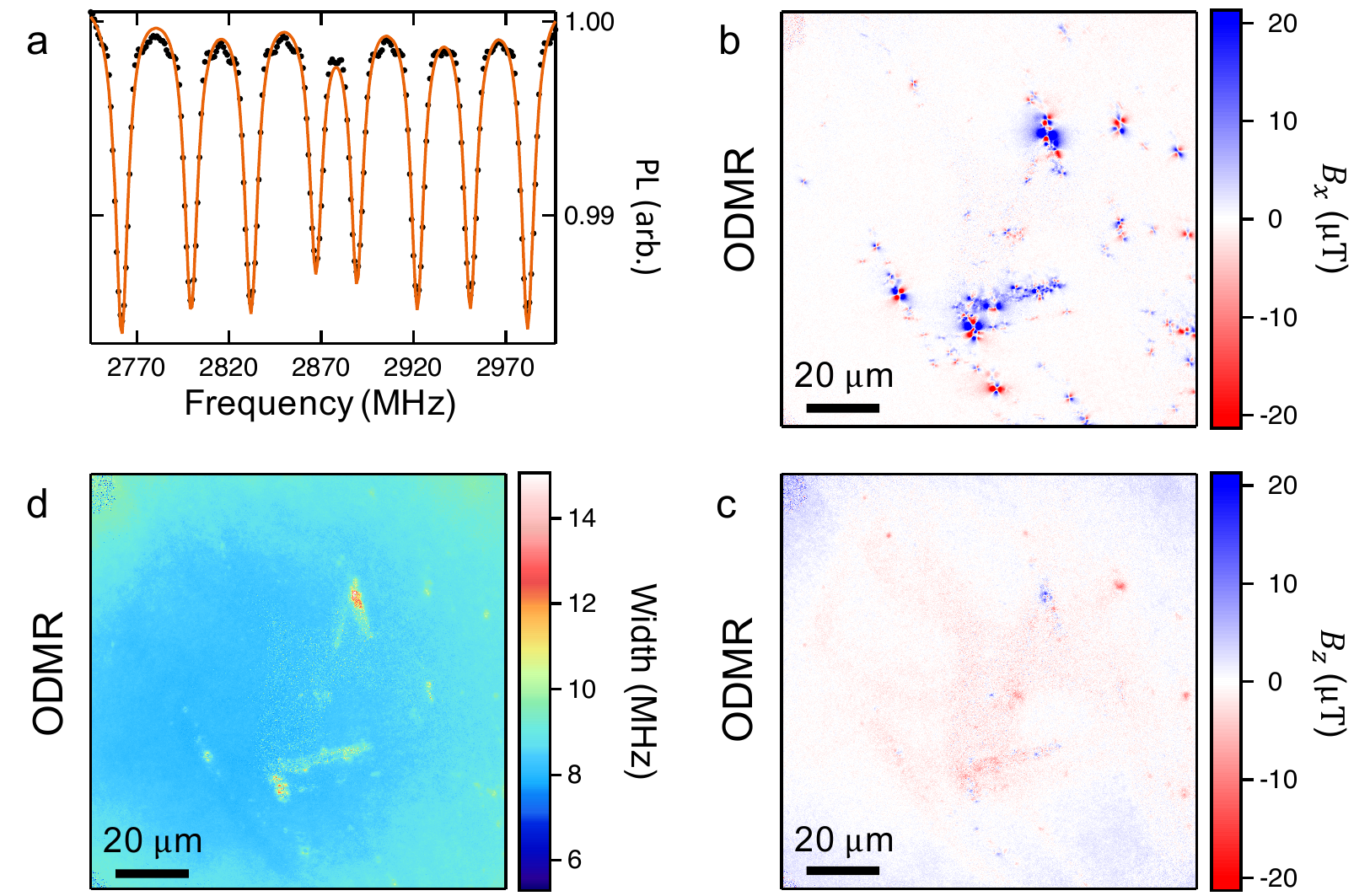}
 		\caption{\textbf{a} ODMR spectrum integrated across the entire inox steel region of interest. \textbf{b} and \textbf{c} show the reconstructed static magnetic field maps in the y-direction, $B_y$, and z-direction, $B_z$, respectively. \textbf{d} Linewidth map of the lowest frequency NV spin-transition. This was the same transition for which the optical contrast was mapped in Fig. 2d.}
 		\label{FigS3}
 	\end{center}
\end{figure}

The ODMR spectrum integrated across the entire inox steel region of interest is shown in Fig. \ref{FigS3}, showing the eight clearly resolved transitions from the four NV families. The magnetic field map, reconstructed from pixel-by-pixel fitting of the transition frequencies, in the x- and z-directions, are shown in Figs. \ref{FigS3}b and c. The static field in the x-direction, $B_x$, closely resembles that in the y-direction, $B_y$, shown in the main text. The field in the z-direction, however, is reduced significantly in magnitude as compared to the in-plane counterparts. This effect arises from the close proximity of the sensing layer to the ferromagnetic nanoparticles, where the strong fields seen by the closest NV centres result in them being neglected by the measurements and fitting routines, and hence, the signal is dominated by those seeing lesser field strengths within the same optically resolvable sensing volume \cite{Tetienne2018c}.
 
In the main text, the ODMR contrast map was used to infer a more complex interaction between the NV-layer and deposited material. The reduced contrast seen under the deposited area could result from a broadening of the peaks due to the deposited material, through a static spread in magnetic field, or magnetic fluctuations. However, mapping the width of a single peak, again the lower frequency transition, we see that the width is only significantly increased due to the fields arising from the ferromagnetic nanoparticles, an expected result \cite{Tetienne2018c}. This implies that the reduced ODMR contrast in the abraded regions is not directly caused by the deposited material, and must be instead explained by a more subtle effect whereby the photo-dynamics of the NV is affected in a way that results, for instance, in a reduced spin initialisation.
 
\section{EDXS of inox steel deposition} \label{EDXS}
 
One of the outstanding challenges in working with atomically thin samples on a rough diamond surface, is characterising the deposited materials independently to the NV measurements. Raman spectroscopy was attempted on similar samples to those presented in the main text, however, their small contribution to the total sensing volume and the background fluorescence of the NV layer make this a particularly difficult measurement. Energy-dispersive X-ray spectroscopy (EDXS) is a useful technique for characterising such samples, as it combines high-resolution SEM imaging, which typically shows contrast between bare substrates and monolayer coverage, with spectroscopy that can identify the atomic composition of larger particles ($>100$\,nm). Here we present results for an inox steel deposited sampled, similarly prepared to that presented in the main text.
 
\begin{figure}[t]
 	\begin{center}
 		\includegraphics[width=0.4\textwidth]{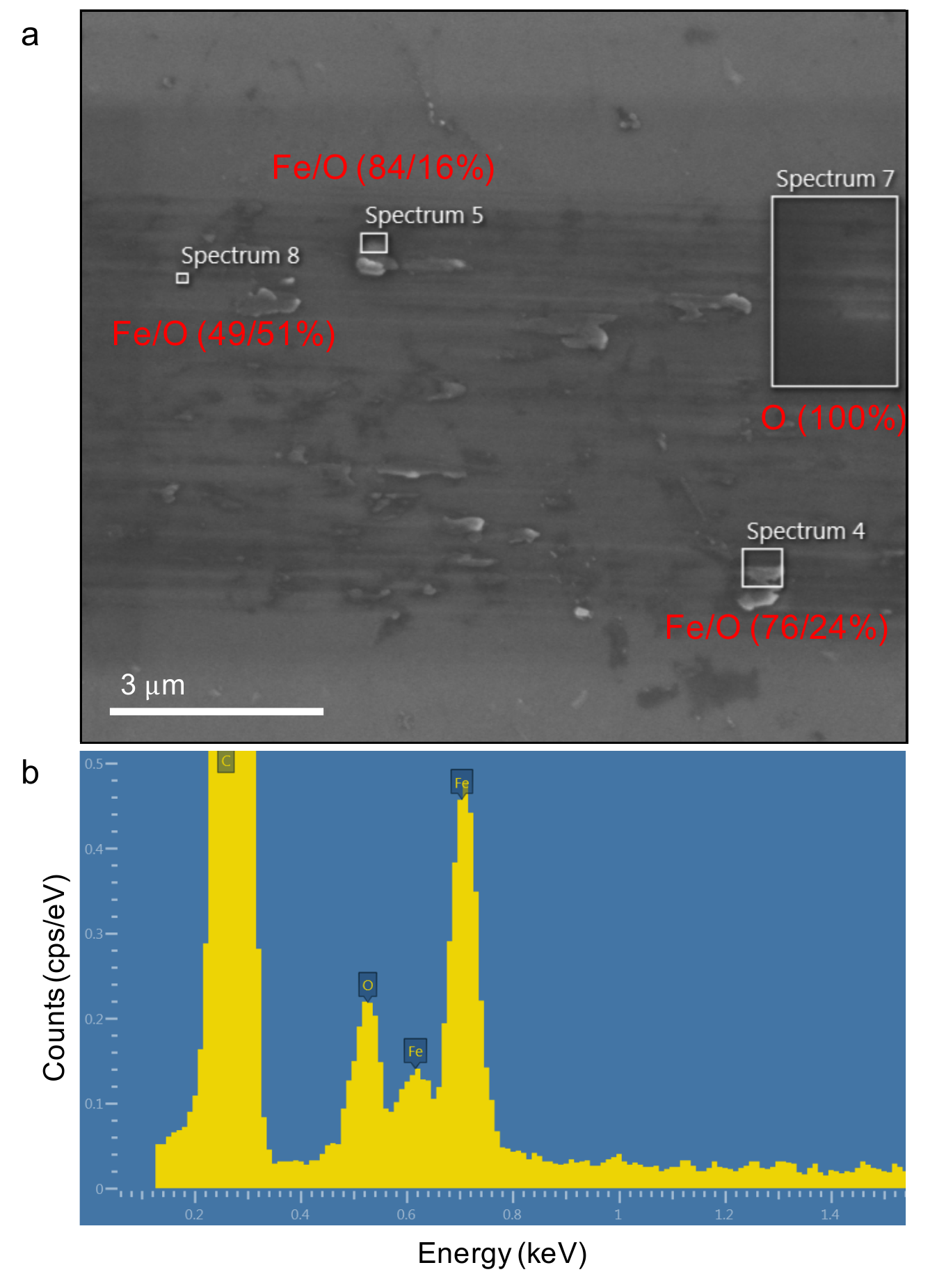}
 		\caption{\textbf{a} SEM image of a inox steel deposited region (dark streak), annotated with results from EDXS  giving relative atomic abundances, excluding the dominant contribution of carbon from the diamond substrate. \textbf{b} Representative EDXS spectra of the one of the large particles deposited within the streak.}
 		\label{FigS4}
 	\end{center}
\end{figure}
 
Figure \ref{FigS4}a shows an SEM image of the deposited region, showing features similar to those seen in the SEM imaging of the sample presented in the main text. A banded streak, approximately $5$\,$\mu$m wide, is seen with larger particles embedded in it. The EDXS spectra of these particles show the presence of Fe and O in abundances commensurate with Fe$_2$O$_3$ and Fe$_3$O$_4$ (labels on Fig. \ref{FigS4} give the atomic abundances excluding the dominant C peak from the substrate). A representative spectrum of these particles is shown in Fig. \ref{FigS4}b, where the large peak at low energy is characteristic of carbon. It is important to note that spectrum 7 in Fig. \ref{FigS4}a, acquired from a region of the streak free of large particles, shows only the presence of oxygen, likely due to the surface termination of the diamond, again suggesting a sparse coverage of material from the abrasive deposition.
 
\section{Comparison of abrasively deposited materials} \label{Materials}
 
 In the main text, PL, $T_1$ and AFM data was presented for a range of materials, to compare the consequences of their deposition on diamond. The NV imaging data presented was restricted to regions for which spatially correlated AFM imaging was achieved, however, all imaging was acquired over an approximately $100$\,$\mu$m field of view. Here we present the full field of view data for the PL and $T_1$ imaging for completeness. 
 
\begin{figure*}[t]
 	\begin{center}
 		\includegraphics[width=1.0\textwidth]{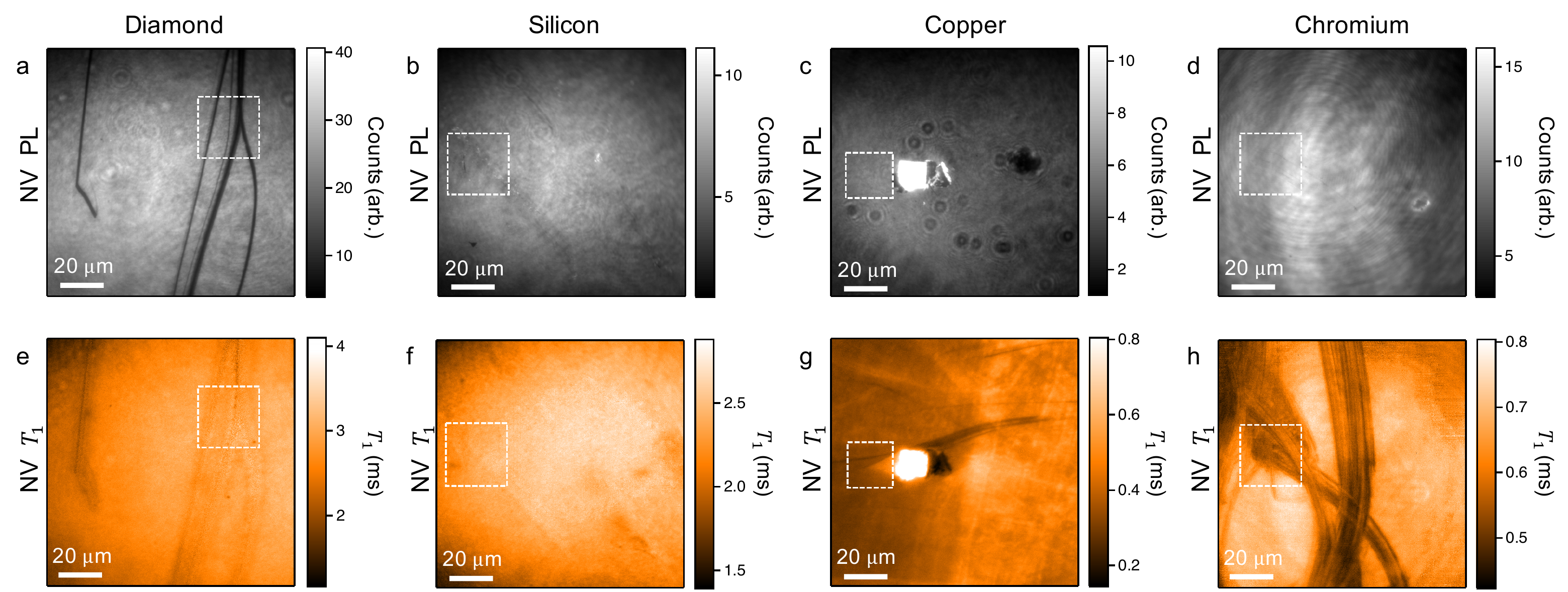}
 		\caption{\textbf{a} - \textbf{d} PL and \textbf{e} - \textbf{h} $T_1$ images across the full field of view for the diamond scribe, silicon wafer, copper, and chromium abrasive depositions respectively. The regions presented in Fig. \ref{Fig3} are highlighted (white boxes).}
 		\label{FigS5}
 	\end{center}
\end{figure*}
 
Figure \ref{FigS5} shows the full field of view PL and $T_1$ images of the sample regions scratched with a diamond scribe, silicon wafer, copper, and chromium. The scribe region shows a large cavitation pattern in PL [Fig. \ref{FigS5}a], due to the removal of NV centres and the consequent destabilisation of the NV charge as discussed, however, the $T_1$ is largely uncharged [Fig. \ref{FigS5}e]. Similarly, the silicon wafer region shows a clear band of quenched PL [Fig. \ref{FigS5}b], a small region of which was presented in the main text, with bright centres arising from larger Si particles ($>1$\,$\mu$m), but again, no significant change in $T_1$. Background characterisation of the copper region, which was centered around a growth pyramid [Fig. \ref{FigS5}], revealed a pre-existing $T_1$ boundary, where the left hand side of the imaged region has a slightly shorter relaxation time than the right. This feature remained after the copper deposition, which did not directly affect PL [Fig. \ref{FigS5}c], but quenched the $T_1$ [Fig. \ref{FigS5}g]. The region for which correlated AFM was achieved lay to the left hand side of the pyramid. The large feature seen to the right hand side of the pyramid is a large chunk of metallic copper, which shows bright in PL, and strongly quenches $T_1$. No discernible step height was measured across the steak of quenched $T_1$ running across the centre of the image. Finally, the chromium deposited region shows only a slight PL quenching parallel to the scratch direction as discussed in the main text [Fig. \ref{FigS5}d], whereas the $T_1$ quenching is significant [Fig. \ref{FigS5}h].
 
\begin{figure*}[t]
 	\begin{center}
 		\includegraphics[width=1.0\textwidth]{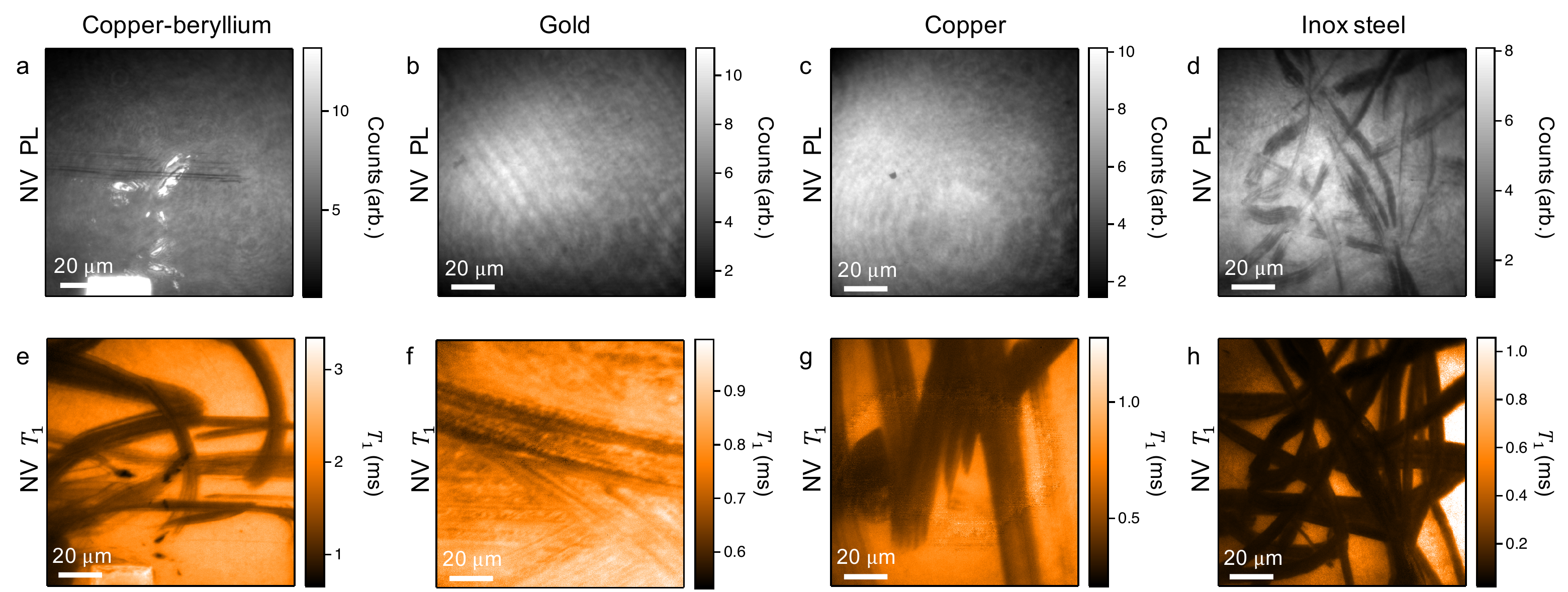}
 		\caption{\textbf{a} - \textbf{d} PL and \textbf{e} - \textbf{h} $T_1$ imaging of additional copper-beryllium, gold, copper, and inox steel abrasive depositions respectively. For Fig. \textbf{h} ,the z-scale is capped at $1.05$\,ms to emphasise contrast in the left hand side of the image. The $T_1$ time is as up to $2$\,ms long in the right hand side of the image}
 		\label{FigS6}
 	\end{center}
\end{figure*}
 
In addition to the materials discussed above for which spatially correlated AFM data was collected, additional scratches were made with a range of secondary materials, for which only NV data was collected. The PL and $T_1$ images of regions scratched with a copper-beryllium composite, gold, copper, and inox steel are shown in Figs. \ref{FigS6}a - d and Figs. \ref{FigS6}e - h respectively. None of the materials show particularly strong PL quenching, with the exception of inox steel which shows a stronger quenching than shown in the main text (the bright regions seen for the copper-beryllium scratches as due to large chunks of the composite sitting on the surface, and the dark lines running horizontally across the field of view were pre-existing scratches in the sample [Fig.\ref{FigS6}a]) but all result in clear quenching of the NV-layer $T_1$ relaxation time. Here the total added noise for each deposition is $0.6$\,kHz, $0.7$\,kHz, $2.0$\,kHz, and $5.0$ kHz  for the copper-beryllium, gold, copper, and inox steel respectively. The added noise was calculated by comparing the $T_1$ times under and beside the abraded regions for the copper-beryllium, copper, and inox steel, where the gold compared to a background measurement, as the $T_1$ decreased across the entire field of view after the scratch due to the blunt nature of the macroscopic block used. The result for gold is particularly interesting given it lacks a candidate form for the spin-noise origin, suggesting the effect is likely an indirect effect one on the diamond itself. We note that the gold abrasion was performed on a polished diamond surface, which is responsible for the grating (polishing marks) visible in Fig. \ref{FigS6}c.
 
\section{Johnson-Nyquist noise from metallic thin films} \label{JohnsonCalc}
 
Ref. \cite{Ariyaratne2017} derive a model for the relaxation rate induce by Johnson-Nyquist noise arising from a metallic film as
 
\begin{equation}
 \Gamma_{\text{metal}} = \gamma^2 \dfrac{\mu_0^2 k_B T \sigma}{8 \pi} \left(\dfrac{1}{d} - \dfrac{1}{d + t_{\text{film}}} \right)
\end{equation}
 
\noindent where $\gamma$ is the electron gyromagnetic ratio, $\mu_0$ is the vacuum permeability, $T$ is temperature, $\sigma$ is the metal conductivity, $d$ is the separation between the sensing layer and the metallic film, which has thickness $t_\text{film}$. Calculating the expected noise for a film thickness of $1$\,nm with the conductivity of bulk copper, $\sigma = 6.0\times10^7$\,$\Omega^{-1}$m$^{-1}$, gives a noise $\Gamma_{\text{metal}}=2.0$\,kHz. However, once the reduced conductivity at film thicknesses less than the mean free path are taken into account, which, conservatively, reduces conductivity by an order of magnitude at a $5$\,nm film thickness \cite{Lacy2011}, the calculated noise is well below that observed for even the lower conductivity metals that were abrasively deposited. Additionally, this model does not account for the lateral confinement of our particulate deposition, which will only further reduce the conductivity \cite{Ariyaratne2017}.
 
\section{Double electron-electron resonance imaging} \label{DEER}
 
 \begin{figure}[t!]
 	\begin{center}
 		\includegraphics[width=0.5\textwidth]{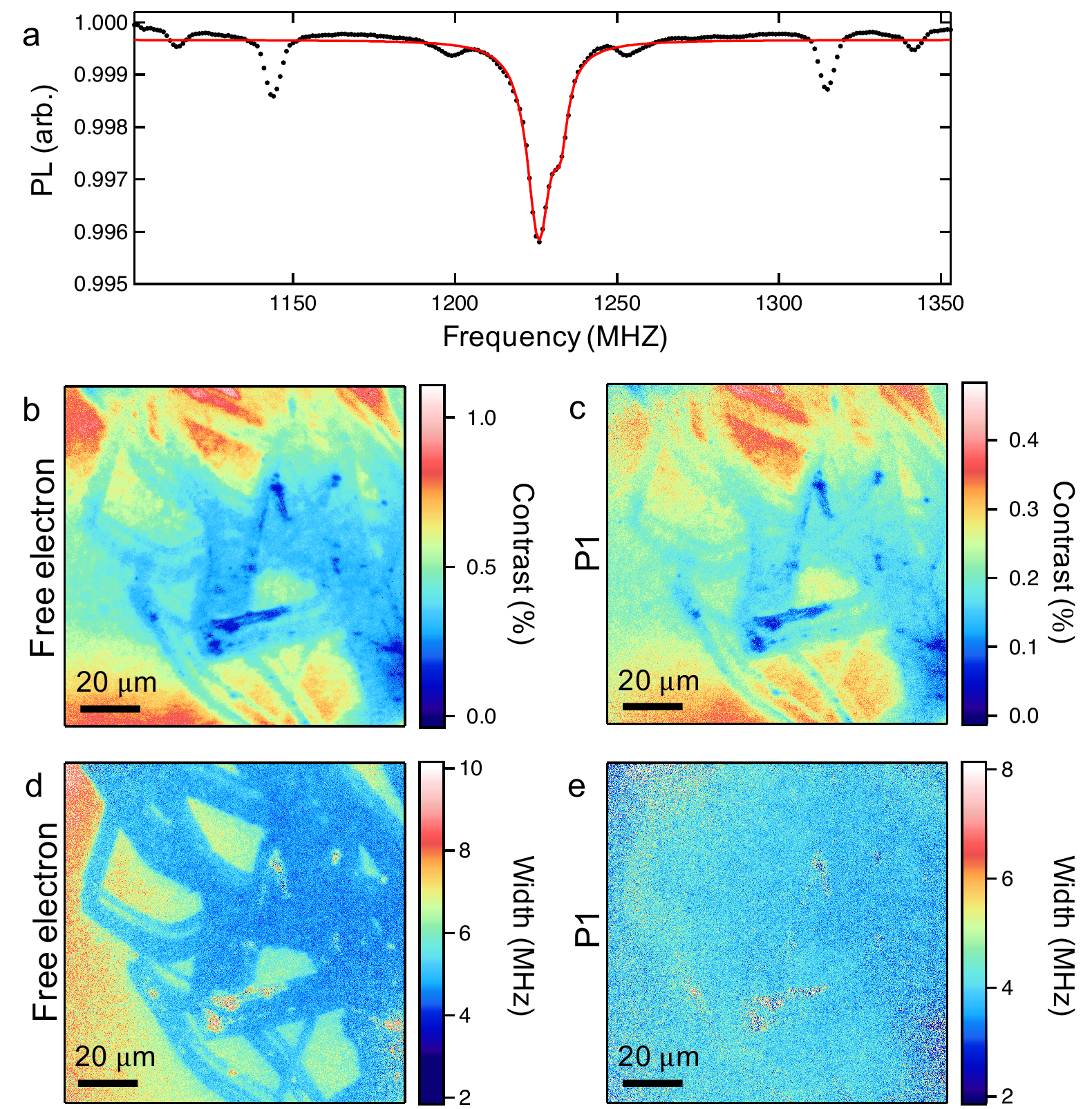}
 		\caption{\textbf{a} DEER spectrum centred around the free-electron peak at $437$\,G. The large central peak is associated with free-electron spins. The shoulder and five outer peaks are associated the electron spin of substitutional nitrogen (P$1$) centres intrinsic to the diamond. \textbf{b} and \textbf{c} DEER spectrum peak contrast for the central free-electron, and P$1$ shoulder peaks respectively. \textbf{d} and \textbf{e} Width maps of the free-electron and central P$1$ peaks respectively.}
 		\label{FigS7}
 	\end{center}
\end{figure}

Double electron-electron resonance (DEER) is a standard technique in NV sensing used to detect electron spin species \cite{Grotz2011,Mamin2012}. Target species within some frequency range are driven synchronously with a spin-echo sequence driven on one family of the NV layer, resulting in a detectable phase accumulation. Figure \ref{FigS7}a shows the DEER spectrum integrated across the entire inox steel deposited region (identical to the region shown in Figs. $1$ and $2$), taken at $437$\,G. Here we fit the frequency of the free-electron peak (centre), with a fixed relative shift of the substitutional nitrogen (P$1$) peak, given by the transverse hyperfine coupling parameters with the $^{14}$N nucleus \cite{Wood2016}, and produce maps of the contrast [Figs. \ref{FigS7}b and c] and width [Figs. \ref{FigS7}d and e] of each peak.
 
The optical contrast of both the free-electron peak and the central P$1$ centre peak [Figs. \ref{FigS7}b and c] closely resembles the optical contrast map of the NV ODMR measurement, which again suggests that the charge stability of the NV centre are affected during the dark time of the measurement. Comparing the widths of the two peaks across the field of view, both are broadened at the sites of ferromagnetism, as for the NV ODMR, but the free-electron peak is narrowed under the abraded region. The reduced contrast and width of the free-electron peak suggest a reduced visibility of these species to the DEER sequence under the abraded regions, which may result from a reduction in their own $T_1$ relaxation times, or changes to their population, again suggestive of an effect from the deposited material on the NV environment.
 
\bibliographystyle{apsrev4-1}
\bibliography{ScratchPaper}
	   
\end{document}